\documentclass[final,doublecol]{epl2}

\usepackage{graphics,graphicx}
\usepackage{amsmath}
\usepackage{amssymb}

\usepackage[T1]{fontenc}

\title{Remanence effects in the electrical resistivity of spin glasses}

\author{Thibaut Capron\inst{1}, Angela Perrat-Mabilon\inst{2}, Christophe Peaucelle\inst{2}, Tristan Meunier\inst{1}, David Carpentier\inst{3}, Laurent~P.~L\'{e}vy\inst{1,4,5}, Christopher B\"{a}uerle\inst{1} and Laurent Saminadayar\inst{1,4,5,*}}
\shortauthor{T. Capron \etal}

\institute{
\inst{1}Institut N\'{e}el, B.P. 166, 38042 Grenoble Cedex 09, France \\
\inst{2}Institut de Physique Nucl\'{e}aire de Lyon, Universit\'{e} de Lyon, Universit\'{e} de Lyon1, \textsc{cnrs} \& \textsc{in2p3}, 04 rue Enrico Fermi, 69622 Villeurbanne Cedex, France \\
\inst{3}Laboratoire de Physique, \'{E}cole Normale Sup\'{e}rieure de Lyon, 47 all\'{e}e d'Italie, 69007 Lyon, France \\
\inst{4}Universit\'{e} Joseph Fourier, B.P. 53, 38041 Grenoble Cedex 09, France \\
\inst{5}Institut Universitaire de France, 103 boulevard Saint-Michel, 75005 Paris, France 
}

\date{\today}

\pacs{73.23.-b}{}
\pacs{03.65.Bz}{}
\pacs{75.20.Hr}{}
\pacs{72.70.+m}{}
\pacs{73.63.-b}{}
\pacs{75.10.nr}{}

\abstract{
We have measured the low temperature electrical resistivity of $Ag:Mn$ mesoscopic spin glasses prepared by ion implantation with a concentration of $700\, ppm$. As expected, we observe a clear maximum in the resistivity $\rho (T)$ at a temperature in good agreement with theoretical predictions. Moreover, we observe \emph{remanence effects} at very weak magnetic fields 
for the resistivity below the freezing temperature $T_{sg}$: upon Field Cooling (\textsc{fc}), we observe clear deviations of $\rho (T)$ as compared with the Zero Field Cooling (\textsc{zfc}); such deviations appear even for very small magnetic fields, typically in the Gauss range. This onset of remanence for very weak magnetic fields is reminiscent of the typical  signature  on magnetic susceptibility measurements of the spin glass transition for this generic glassy system.
}

\begin{document}

\maketitle


Spin glasses are one of the most intriguing states of matter, and have been the subject of intense theoretical as well as experimental works over the last decades~\cite{Mydosh_Book,Fisher_Book}; this is due to the fact that they can be considered as the most generic example of a glassy phase~\cite{Les_Houches_Book} but also because the very nature of the ground state of these systems is still heavily debated and may eventually consist in a very peculiar state of matter~\cite{Mezard_84_AA,Fisher_86_AA}. In spin glasses (\textsc{sg}), magnetic moments occupying random positions in a metallic host interact \textsl{via} \textsc{rkky} interactions; the randomness in the position of these moments leads to a randomness in both the sign and the amplitude of their mutual interaction. Below the spin glass transition temperature $T_{sg}$, the interplay between disorder and frustration will freeze the magnetic spins in a complex way far from being well understood. Most of the experimental studies on spin glasses has been carried out on macroscopic systems: only few papers addressed the question of \emph{mesoscopic spin glasses}, either theoretically~\cite{Altshuler_85_AA,Feng_87_AA,Carpentier_08_AA} or experimentally~\cite{Israeloff_89_AA,Vegvar_91_AA,Jaroszynski_98_AA}. Moreover, most of the measurements concerned thermodynamic quantities, mainly magnetic susceptibility~\cite{Cannella_72_AA,Nagata_79_AA,Mulder_81_AA}; very few experiments addressed the question of the \emph{transport properties} of mesoscopic spin glasses~\cite{Tindal_74_AA,Ford_76_AA,Campbell_82_AA} and their interpretation remains a theoretical challenge~\cite{Larsen_76_AA,Sheikh_89_AA,Vavilov_03_AB}.

The most convincing signature that the spin glass transition may actually be a real phase transition is the cusp which appears in the low field magnetic susceptibility at the transition temperature~\cite{Cannella_72_AA}. On the other hand, one of the most fascinating properties of spin glasses, which makes this state of matter a very peculiar one,  is the onset of \emph{remanence effects for very weak magnetic fields} below $T_{sg}$: the dc susceptibility $\chi_{dc}$ of a spin glass depends strongly on the way the experiment is performed, in particular if the cooling through $T_{sg}$ has been made under a small magnetic field (Field Cooled) or without magnetic field (Zero Field Cooled). Experimentally, $\chi_{dc}$ is much larger after Field Cooling than after Zero Field Cooling~\cite{Cannella_72_AA, Djurberg_99}. We would like to stress that the magnetic fields involved in these experiments are very small, typically in the Gauss range, i.e. $\mu_{B}B\ll k_{B}T_{sg}$, with $\mu_{B}$ the Bohr magneton, $k_{B}$ the Boltzmann constant and $B$ the magnetic field: this means that the remanence effect has nothing to do with a trivial effect of polarization of the spins, but is a true signature of the somehow \textquotedblleft unconventional\textquotedblright~nature of the spin glass transition.

In this Letter, we show that such remanence effects can also be observed on the \emph{electrical resistivity} of metallic spin glasses. We show that the resistivity $\rho$ as a function of temperature $T$ of a $Ag:Mn$ spin glass exhibits a clear maximum at a temperature, in excellent agreement with the theoretical prediction of Vavilov \textsl{et al.}~\cite{Vavilov_03_AB} and with previous experiments on bulk spin glasses~\cite{Tindal_74_AA,Ford_76_AA,Campbell_82_AA}. In addition, we observe a remanence effect on the $\rho(T)$ curve, an effect that has never been observed so far on transport properties: when cooling down from above $T_{sg}$ under \emph{very small} magnetic field (typically in the Gauss range), i.e. \emph{when Field Cooling the sample, the resistivity is significantly lower compared to when cooling down under zero magnetic field}. Such irreversibilities disappear at the transition temperature, as observed on thermodynamic quantities.

\begin{figure}[th!]
\includegraphics[width=8.5cm]{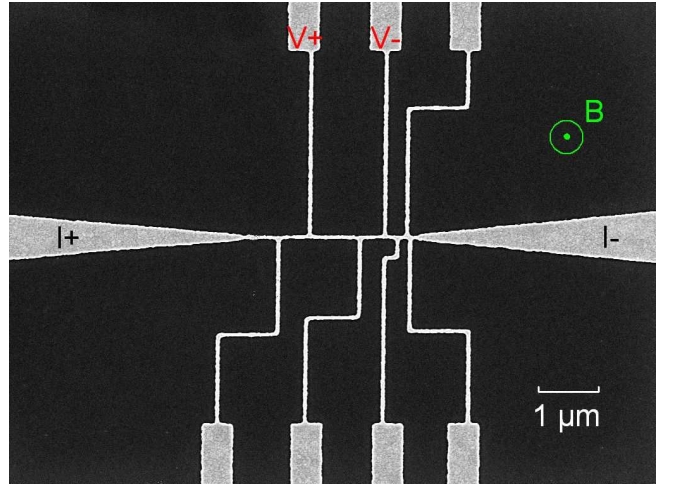}
\caption{(Color online) Scanning Electron Microscope (\textsc{sem}) image of the sample used in the experiment. Current and voltage probes are indicated, and the magnetic field is applied perpendicularly to the plane. The distance between the two voltage probes is $\approx 1\,\mu m$.}
\label{Sample}
\end{figure}

Samples are fabricated on a silicon substrate using electron-beam lithography on polymethyl-methacrylate resist. Silver has been evaporated in an electron gun evaporator from a $99.9999\,\%$ purity source without adhesion-layer. The geometry of the samples can be seen on figure~\ref{Sample}: it consists of a short wire of length $\approx 1\,\mu m$, width $\approx 50\, nm$ and thickness $\approx 50\, nm$. Several voltage contacts have been designed in order to perform four-points measurements. $Mn^{2+}$ ions are then implanted at an energy of $70\, keV$; the energy has been chosen so that the implantation takes place mainly in the silver layer. The reason for choosing Manganese as magnetic impurity is two-fold: firstly, $Ag:Mn$ is a \textquotedblleft generic\textquotedblright~spin glass system which has been widely studied~\cite{Tindal_74_AA,Ford_76_AA,Campbell_82_AA} and thus its physical characteristics are well known. Secondly, the Kondo temperature $T_K$ of manganese in silver is in the milliKelvin range~\cite{Rizzuto_74_AA}: it is thus easy to reach a regime where the spin glass transition appears at a much higher temperature than the Kondo temperature, $T_{sg}\gg T_{K}$; in this case the Kondo-related physics can be neglected~\cite{Mallet_06_AA}. In our case the ion dose, measured \textit{via} the current out of the implanter, has been chosen such that the final concentration in the sample is $700\, ppm$; this leads to a spin glass transition temperature of $\approx 700\, mK$, i.e. $T_{sg}> 10\,T_{K}$. Finally, let us stress that the implantation process excludes the possibility of clustering or diffusion of the $Mn^{2+}$ ions as no further annealing has been performed on the samples.

Resistance measurements have been performed using a standard $ac$ lock-in technique with a very low current ($\lesssim 500\, nA$) at a frequency of $11\, Hz$. The signal is amplified using an ultra-low noise (${400\, pV}/{\sqrt{Hz}}$) home made preamplifier at room temperature. As the relative variations of the signal we want to extract are very small, the measurements are made in a bridge configuration in order to compensate for the residual resistance of the sample. The sample is fixed at the coldest part of a dilution refrigerator and connected to the measurement setup \textit{via} coaxial cables which ensure a very efficient radio-frequency filtering~\cite{Glattli_97_AA} and thus allow electrons to be cooled down to less than $40\, mK$~\cite{Bauerle_05_AA}. At $4.2\, K$, the resistance of our samples is of the order of $\approx 100\,\Omega$.

\begin{figure}[th!]
\includegraphics[width=8.5cm]{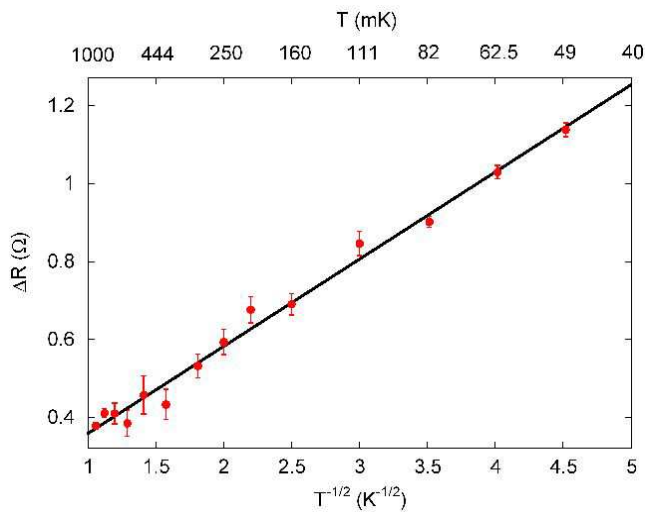}
\caption{(Color online) Resistance as a function of $1/\sqrt{T}$ for a pure silver wire. The variation is perfectly linear and the fit allows to determine the coefficient $\alpha_{e-e}$ of the equation~\ref{AA}.}
\label{Fig2}
\end{figure}

At low temperature, the variation of the resistance as a function of temperature is due to three main contributions: the electron-phonon interaction, mainly efficient at \textquotedblleft high\textquotedblright~temperature, the electron-electron interaction which leads to a modification of the density of states at the Fermi energy (the \textquotedblleft Altshuler-Aronov\textquotedblright~(\textsc{aa}) correction)~\cite{Akkermans_07_AA} and an extra term $\Delta R_{sg}$ which we attribute to the glassy nature of the system:
\begin{equation}
\Delta R=\Delta R_{e-ph}+\Delta R_{e-e}+\Delta R_{sg}
\label{R_T}
\end{equation}
$\Delta R_{e-ph}$ being dominant only at high temperature ($T > 4\, K$). The second one can be expressed as~\cite{Akkermans_07_AA}:
\begin{equation}
\Delta R_{e-e}=\alpha_{e-e}\frac{R^{2}}{R_K}\frac{L_T}{L},
\label{AA}
\end{equation}
where $R_K$ is the quantum of resistance $R_K=h/e^2$ with $h$ the Planck's constant and $e$ the charge of the electron, and $L_T$ is the thermal length, $L_T=\sqrt{\frac{\hbar D}{k_{B}T}}$, $D$ being the diffusion coefficient, $k_{B}$ the Boltzmann's constant and $T$ the temperature. In order to evaluate the prefactor $\alpha_{e-e}$, we have measured the \textsc{aa} correction to the resistance on a longer and pure (non implanted) wire fabricated \emph{in the same evaporation run} as the short wires: when plotted as a function of $1/\sqrt{T}$, the low-temperature resistance of the wire varies linearly, as can be seen on figure~\ref{Fig2}; this ensures a good and reliable determination of the parameter $\alpha_{e-e}$, that we find equal to $2.42$, in good agreement with values found in the literature~\cite{Saminadayar_07_AA}.

\begin{figure}[th!]
\includegraphics[width=8.5cm]{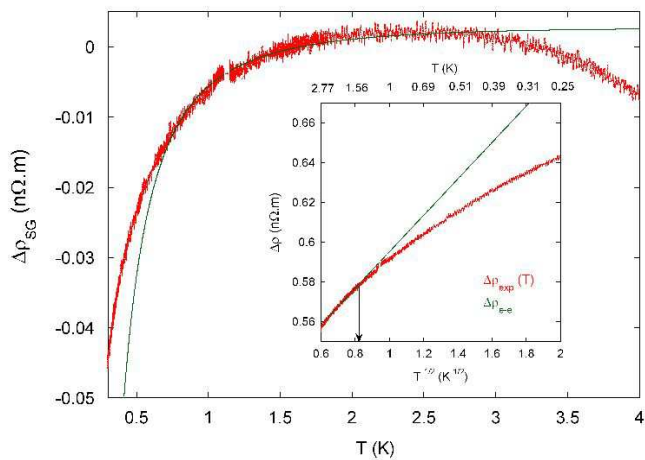}
\caption{(Color online) Resistivity as a function of temperature for an $Ag:Mn$ spin glass at a concentration of $700\, ppm$, after subtraction of the Altshuler-Aronov electron-electron contribution to the resistivity. Inset: raw data before subtraction of the \textsc{aa} contribution.}
\label{Fig3}
\end{figure}

We now turn to the resistivity measurements on the $Ag:Mn$ spin glass system under zero magnetic field. Resistivity as a function of $1/\sqrt{T}$ is displayed on the inset of figure~\ref{Fig3}: contrary to the case of pure silver, clear deviations from the $1/\sqrt{T}$ are observed below $1.5\, K$, and can be attributed to the spin-glass transition. In order to separate the spin-glass contribution to the resistivity, we have subtracted the electron-electron interaction contribution to the resistivity (second term of the equation~\ref{R_T}); the resulting $\Delta\rho_{sg} (T)$ is displayed on figure~\ref{Fig3}.

A clear maximum in the $\Delta\rho_{sg} (T)$ curve appears at a temperature of $T_{m} \approx 2\, K$. This maximum, which has been observed in different magnetic alloys at higher concentrations~\cite{Ford_76_AA}, is a clear signature of the spin glass transition. We can go further in the analysis by fitting the data with the formula suggested by Vavilov \textit{et al.}~\cite{Vavilov_03_AB}:
\begin{equation}
\Delta\rho_{sg} (T) \propto \frac{S(S+1)}{{\ln^{2}(T/T_{K})}} \left(1-\frac{\beta T_{sg}}{T}\right),
\label{Delta(rho)}
\end{equation}
where $S$ is the spin of the magnetic impurity, $T_{K}$ its Kondo temperature ($40\, mK$ for $Mn$ in $Ag$), $T_{sg}$ the spin-glass transition temperature and $\beta$ a numerical factor of order $1$. As can be seen on figure~\ref{Fig3}, a qualitative agreement is obtained between formula~\ref{Delta(rho)} and the experimental data using the parameters $T_{sg} \simeq 700\, mK$ and $\beta = 2.33$ as expected for a spin $S=5/2$~\cite{Vavilov_03_AB}. The discrepancies can be understood by noting that the theoretical formula is valid only for $T_{sg} \leq T \leq T_{m}$: within this interval, one can verify that the agreement is quantitatively good. Moreover, the $T_{sg}$ obtained by this procedure is in agreement with the transition temperature measured on macroscopic $Ag:Mn$ spin glasses of equivalent concentration using other measurement techniques (magnetic susceptibility, specific heat)~\cite{Tindal_74_AA,Ford_76_AA}.

One of the most striking characteristics of spin glasses lies in the onset of remanence effects below $T_{sg}$; the most spectacular manifestation of this remanence appears in the static magnetic susceptibility $\chi_{dc}$ which has been shown to strongly depend on the history of the sample~\cite{Nagata_79_AA}: even for very small magnetic field (in the Gauss range), if the field is applied above $T_{sg}$ and kept fixed down to a temperature below $ T_{sg}$ (\textsc{fc}), $\chi^{(\textsc{fc})}_{dc}$ is larger than after Zero-Field Cooling (\textsc{zfc}) and roughly temperature independent \cite{Jonsson_05_AA,Lundgren_85_AA}.

\begin{figure}[th!]
\includegraphics[width=8.5cm]{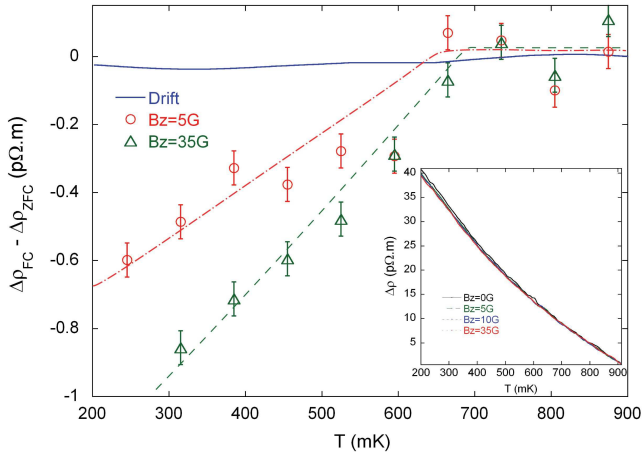}
\caption{(Color online) Resistivity as a function of temperature for the $Ag:Mn$ sample under small magnetic fields of $5\, G$ and $35\, G$, after subtraction of the Zero Field Cooled contribution (data measured under $10\, G$ have been removed for clarity). The solid line represents the typical drift of our experimental setup over the time of the measurement. The cooling rate is $\approx 5\, mK min^{-1}$. Dashed lines are guides to the eyes. Inset: raw data before subtraction of the \textsc{zfc} contribution.}
\label{irreversibility}
\end{figure}

Spin-glass transitions have been previously observed in transport using the anomalous Hall effect~\cite{Vloeberghs_90_AA}, but never in the resistivity.  In particular, remanence effect are well known in the susceptibility but have never been identified in the resistivity. Such a characteristic effect has never been observed on any physical quantity other than the magnetic susceptibility; in particular, no remanence effect has never been reported concerning the transport properties of spin-glasses. We have thus measured the \textsc{zfc} and \textsc{fc} resistivity of our $Ag:Mn$ alloy under small magnetic fields of $5\, G$, $10\, G$ and $35\, G$ (applied above $T_{sg}$ and kept fixed during cooling down). Raw data are presented on the inset of figure~\ref{irreversibility}: the change in the resistivity when decreasing the temperature is mainly due to the electron-electron interaction (\textsc{AA} correction). In order to highlight the onset of remanence on the resistivity of the spin glass below $T_{sg}$, we have subtracted to the Field Cooled ($\Delta\rho^({\textsc{fc}})(T)$) curves the Zero Field Cooled curve ($\Delta\rho^({\textsc{zfc}})(T)$) used as a \textquotedblleft reference\textquotedblright~measurement. The result ($\Delta\rho^({\textsc{fc}})(T) - \Delta\rho^({\textsc{zfc}})(T)$) is displayed on figure~\ref{irreversibility}. In order to estimate the uncertainty on these data, we have also measured the drift of our experimental setup over the typical time of our measurements (several hours), at constant temperature: the result is displayed on the same graph, and clearly shows that this drift is much smaller than our signal; moreover, the \textsc{fc} curve is perfectly reversible, i.e. the $\Delta\rho_{\textsc{fc}}(T)$ curve follows exactly the same trace when subsequently heating the sample up to $T > T_{sg}$. We have restricted our measurements to small fields because i) remanence effects appear for very small fields ii) at higher fields, orbital effects like Universal Conductance Fluctuations (\textsc{ucf}) or weak localisation may become the dominant contribution to the variation of the resistivity. Finally, let us stress that this effect cannot be related to the Hall effect as the measurement is made in a longitudinal configuration, under very low magnetic field, and as the spin-orbit coupling in silver is very weak.

It should be stressed that this remanence effect appears \emph{at the transition temperature $T_{sg}$} which has been determined independently from the dependence of the resisitivity as a function of temperature (see figure~\ref{Delta(rho)}) and from the impurity concentration. Moreover our experiment is limited to very low magnetic field ($B\leq 50\,G$) and thus no orbital or classical effects can account for the observed variation of the resistivity. These two points allow to assert that this effect is indeed due to the magnetic spins present in the sample and related to the peculiar nature of the ground state of the spin glass. Moreover, our data suggest that the amplitude of the effect depends on the applied field which is consistent with recent measurements of the irreversibility of the magnetic susceptibility of spin glasses~\cite{Djurberg_99_AA,Jonsson_05_AA}.

The original effect in magnetic measurements has been interpreted in two different ways for anisotropic (Ising) spin glasses. In the first approach, the Field Cooling procedure selects an optimal \textquotedblleft free energy valley\textquotedblright and the resulting magnetic susceptibility is close to the equilibrium one. On the other hand, in the Zero Field Cooled procedure, the spins are stuck in an \textquotedblleft energy valley\textquotedblright far from the optimal one. This selection of low energy states by a small magnetic field leads naturally to a $\chi^{(\textsc{fc})}$ larger than $\chi^{(\textsc{zfc})}$~\cite{Parisi_07_AA}. Alternatively, this effect has been attributed to the out of equilibrium character of the spins dynamics below the freezing temperature~\cite{Fisher_88_AA,Lundgren_85_AA,Jonsson_05_AA}. Due to the very slow relaxation dynamics below $T_{sg}$, most of the spins are frozen in an out of equilibrium configuration and this leads to a decrease of $\chi^{(\textsc{zfc})}$ with temperature. A weak magnetic field applied during the cooling process leads to a reorientation of spin domains of size larger than $\xi(B) \simeq B^{-\delta}$ with ${-\delta}$ a positive real number~Ê\cite{Fisher_88_AA}. These reorientations increase the magnetic response for large length scales, leading to the larger $\chi^{(\textsc{fc})}$. That this effect can be observed in electrical resistivity measurements is a remarkable confirmation that spin glass physics can be probed by judiciously using charge transport through a mesoscopic sample. Moreover, this type of measurements allows a direct determination of the $T_{g}$ of mesoscopic wires, although thermodynamic measurements on such small systems are impossible. 

In conclusion, we have measured the resistivity of mesoscopic $Ag:Mn$ spin glass wires. The $Mn^{2+}$ magnetic impurities have been implanted in the metal with an energy of $70\, keV$ and a concentration of $700\, ppm$. At low temperature, the resistivity of the spin-glass as a function of the temperature exhibits a clear maximum at a temperature in excellent agreement with theoretical predictions; when lowering further the temperature, the resistivity decreases, a behaviour that is also captured by recent theoretical models. Finally, we observe the onset of \emph{remanence effects} below $T_{sg}$: when cooling down under a very \emph{small} magnetic field (in the $G$ range) (\textsc{fc} procedure), the temperature dependence of the resistivity is significantly different than the one measured when cooling down with no magnetic field (\textsc{zfc} procedure). This \textsc{fc} behaviour is perfectly reversible and appears at the freezing temperature, similarly to what is observed on bulk samples when measuring the magnetic susceptibility. A deeper theoretical comprehension of the spin glass transition is certainly necessary in order to have a quantitative description of these remanence effects.

\acknowledgments
We are indebted to H. Bouchiat and the Quantronics group for the use of their Joule evaporators. We thank G. Paulin, E. Orignac, L. Glazman, B. Spivak and R. Whitney for fruitful discussions. This work has been supported by the French National Agency (\textsc{anr}) in the frame of its \textquotedblleft Programmes Blancs\textquotedblright~(\textsc{MesoGlass} project).

\end{document}